\documentclass{statsoc}

\usepackage[a4paper]{geometry}
\usepackage{graphicx}
\usepackage[textwidth=8em,textsize=small]{todonotes}
\usepackage{amsmath, amssymb}
\usepackage{bm}
\usepackage{mathtools}
\mathtoolsset{showonlyrefs, showmanualtags} 
\usepackage{multirow}
\usepackage{cancel}
\usepackage{scalerel} 
\usepackage[lined, boxed, noend]{algorithm2e} 
\usepackage{tikz} 
\usetikzlibrary{positioning, calc, fit, arrows, backgrounds}
\usepackage{appendix}
\usepackage{url}

\newtheorem{lemma}{Lemma}
\newtheorem{remark}{Remark}

\newcommand{\state}[1]{\mathrm{#1}}
\newcommand{\tx}[2]{[\mathrm{#1}\mathrm{#2}]}
\newcommand{\sidx}[1]{\scaleto{\state{#1}}{0.75ex}}

\newcommand{\nnz}[1]{\mbox{nnz}\left(#1\right)}

\title[SEIR model]{Bayesian inference for high-dimensional discrete-time epidemic models:  spatial dynamics of the UK COVID-19 outbreak}

\author[C P
Jewell {\it et al.}]{Chris P Jewell} \address{ Department of Mathematics and Statistics,
  Lancaster University, Lancaster, LA1 4YW.  U.K.  } \email{c.jewell@lancaster.ac.uk}
\author[A C Hale {\it et al.}]{Alison C Hale} \address{ Department of Mathematics and
  Statistics, Lancaster University, Lancaster, LA1 4YW, U.K.  }
\author[B M Rowlingson {\it et al.}]{Barry S. Rowlingson} \address{Lancaster Medical
  School, Lancaster University, Lancaster, LA1 4YW, UK}
\author{Christopher Suter} \address{Google Research} 
\author[J M Read {\it et al.}]{Jonathan M. Read} \address{Lancaster Medical School, Lancaster University, Lancaster, LA1 4YW, UK} 
\author[G Roberts {\it et al.}]{Gareth Roberts} \address{
  Department of Statistics, University of Warwick, Coventry, CV4 7AL, U.K.  }

\begin{document}

\begin{abstract}
Stochastic epidemic models which incorporate interactions between space and human mobility are a key tool to inform prioritisation of outbreak control to appropriate locations.  However, methods for fitting such models to national-level population data are currently unfit for purpose due to the difficulty of marginalising over high-dimensional, highly-correlated censored epidemiological event data. Here we propose a new Bayesian MCMC approach to inference on a spatially-explicit stochastic SEIR meta-population model, using a suite of novel model-informed Metropolis-Hastings samplers.  We apply this method to UK COVID-19 case data, showing real-time spatial results that were used to inform UK policy during the pandemic.
\end{abstract}


\section{Introduction}
During the COVID-19 pandemic, accurate situational awareness and the ability to project
case numbers forward in time was an important aspect of adaptive infection control policy.
In the UK, as elsewhere, the epidemic was characterised by dramatically fluctuating case
numbers over space and time, with population behaviour interacting with pathogen
transmission to create a complex and highly-variable disease landscape.  Understanding
this variability in case numbers became an important aspect of epidemic management, and quantitative modelling of the drivers of infection transmission enabled metrics such
as the time-varying reproduction number, intrinsic growth rate, and the impact of
intervention strategies to be estimated. 

Epidemic dynamics are largely driven by the characteristics of the population at risk: 
the behavioural determinants of how individuals interact, socioeconomics and spatially-varying population density and environment in which people live \citep{GrBjFin02}.
In particular, the spatial distribution and mobility of a population 
has a marked impact on the pattern of disease transmission, and understanding how
epidemic behaviour responds to individuals' mobility is a key aspect of outbreak control \citep{KeelEtAl01}. At both a local and national level, infectious disease transmission may be assumed to be facilitated by individual-to-individual contact, and therefore understanding how contact behaviour spreads infection is central in designing successful control interventions whilst minimising public disturbance \citep{KeelEam05}.

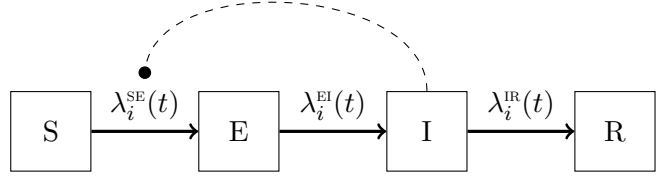
\begin{figure}
  \centering
  \begin{tikzpicture}[auto,
    node distance=40pt,
    box/.style={
      draw=black,
      anchor=center,
      align=center,
      minimum height=30pt,
      minimum width=30pt}]
    \node[box] (S) {$\state{S}$};
    \node[box,right=of S] (E) {$\state{E}$};
    \node[box, right=of E] (I) {$\state{I}$};
    \node[box, right=of I] (R) {$\state{R}$};
    
    \draw[->, very thick] (S) to node (SE) {$\lambda^{\sidx{SE}}_i(t)$} (E);
    \draw[->, very thick] (E) to node (EI) {$\lambda^{\sidx{EI}}_i(t)$} (I);
    \draw[->, very thick] (I) to node (IR) {$\lambda^{\sidx{IR}}_i(t)$} (R);
    \draw[-*,dashed] (I) to [out=90,in=90,looseness=1] (SE);

  \end{tikzpicture}
  \caption{SEIR model showing states (boxes), transitions (arrows), and associated
    transition rates.}
  \label{fig:seir-model}
\end{figure}

Stochastic individual-level state-transition models of epidemics have been extensively used to study the effects of spatial population structure on epidemic spread \citep{KeelEtAl01}.  They are a special case of the more general class of state-transition processes, which consider individuals as transitioning between a number of discrete, mutually-exclusive states reflecting the expected natural history of infection according to a stochastic process \citep{Bar64}.  For example, the classic SEIR model (Figure \ref{fig:seir-model}) assumes individuals begin as \emph{susceptible} to infection, and transition sequentially to \emph{exposed} (i.e. infected but not yet infectious), \emph{infectious}, and finally \emph{removed} (i.e. dead or recovered with immunity from further infection).  Of particular interest is the infection rate, or hazard rate of transitioning from \emph{susceptible} to \emph{exposed}, which given a suitable model can provide valuable insights into how interactions between individuals modulate the propensity for disease transmission as a function of spatial separation, as well as how individual-level covariates affect disease susceptibility and infectivity \citep[e.g.][]{JewEtAl09c, SmEtAl11}.  These models are particularly popular due to their interpretability and compatibility with decision-making questions \citep[e.g.][]{TilEtAl06, PalSilGr2022}.  

Though powerful, individual-level models are computationally intensive, with fast forward simulation and inference methods requiring model-specific optimisations, and inference methods requiring application-specific approximations to allow fast computation of likelihood functions \citep{DearEtAl10, BrEtAl15, SelEtAl18, PrEtAl18}.  Additionally, whilst such methods have been used successfully for population sizes up to approximately 200,000 (commonly suiting livestock disease applications as in \citet{PrEtAl18}), individual-level covariate data is rarely available to support modelling at the individual level for national and international human populations.  In such situations, the population may be stratified by group-level covariates of interest.  Stratification by space leads to the metapopulation model, in which individuals are aggregated into a number of discrete spatial units, with individuals within a spatial unit sharing a set of common covariates \citep{Lev69}.  The exchangability of individuals within each metapopulation hence offers performance gains in both computation time and storage.  

For models with modest numbers of metapopulations (low-dimensional models), a popular
approach is to represent the state transition model as a system of ordinary differential
equations, using a continuous mean-field approximation to the discrete space of numbers of
individuals transiting through the state-transition graph \citep[e.g.][]{LipAlBen21}.
However, increasing numbers of metapopulations (high-dimensional models) divides the
population into smaller units, such that the continuous state-space approximation of the ODE system fails.  In this case, a stochastic implementation is required that captures the random nature of infection transmission among small numbers of individuals within the metapopulation.  Whilst metapopulation models can be implemented as a continuous time-inhomogeneous Poisson process \citep[e.g.][]{MinEtAl11}, a more computationally convenient setup is the discrete-time ``chain-binomial'' model \citep{ becker1981}.  This has the additional computational performance advantage that complexity may be controlled by packaging transition events into time-quanta, rather than having to compute for each individual transition event in continuous time \citep{DiekOtPlBoot21}. 

In a real-time outbreak response context, the utility of epidemic models is maximised through principled parameter inference, providing important policy information through improved situational awareness and forecasting of the ongoing epidemic in space and time \citep{JewEtAl09c, BirEtAl21}.  Given complete epidemic information, i.e. the epidemiological states and transition times for each individual over the course of the epidemic, the likelihood function of a given state-transition model is tractable and amenable to conventional inference methods.  However, in all real-world situations inference is complicated by the inability to directly observe individuals' epidemiological states, or certain transition events.  For example, whereas the times of individual symptom onset or recovery might be well-recorded, infection events are typically unobserved \citep{ChapEtAl18}.  The demands on inference methods for real-time epidemic model fitting therefore present a considerable statistical challenge: not only must the method be rapid in its own right, but must \emph{also} be flexible enough to marginalise over potentially many censored transition events and epidemiological states \citep{SwEtAl22}.

In this paper we shall adopt a Bayesian approach to inference for stochastic metapopulation models, as demonstrated by our COVID-19 application. This has many advantages: it provides a principled and coherent calculus for the measurement of uncertainty both in parameters and predictions of future epidemic scenarios; it gives a natural way of taking into account missing and censored data using data augmentation; it provides a natural framework for  experts to input knowledge to inform inference (via prior elicitation); and finally its implementation is facilitated through powerful MCMC techniques. That being said, MCMC methods are often seriously challenged by spatial applications at scale due to the strong correlations inherently present within such models.  For individual-level models, data-augmentation methodology has become state-of-the-art for unbiased parameter estimation, allowing efficient marginalisation over the censored data \citep{NealRob04, JewEtAl09c, PrEtAl18, ChapEtAl18}.  However, these methods scale poorly to large population sizes, and are therefore unsuitable for the COVID-19 epidemic at hand.  Whilst particle filter methods are beginning to show promise for fitting larger-scale epidemic models, they are inherently constrained by poor performance as the dimensionality of the population stratification increases \citep{RimJewFear23}.  Inspired by existing data-augmentation methodology, we therefore develop a novel
MCMC method suited to metapopulation models, and which are amenable to rapid implementation through efficient multi-core computing.  Our innovations allow us to analyse national-scale outbreaks in a timely manner, capturing spatiotemporal dependence between disease prevalence and incidence.  This then provides estimates of quantities such as the degree of interaction between population strata and reproduction number, and predictions of the ongoing disease trajectory.

Between May 2020 and March 2022, the modelling approach to the
UK's COVID-19 outbreak we describe in this paper was used to provide information to the UK
government, via the Scientific Pandemic Influenza Group on Modelling, Operational
sub-group (SPI-M-O) of the Scientific Advisory Group on Emergencies (SAGE).

\subsection{Spatial COVID-19 cases in the UK}\label{sec:motivating-dataset}

Data on the daily number of new cases of COVID-19 in each of 382 Local Authority
Districts (LADs) in the UK is available from the UK Government Coronavirus website
\citep{coronaviruswebsite, coviddatadownload}.  This dataset contains the number of people
$y_{it}$ who submitted a positive test sample for COVID-19 on day $t$ in LAD $i$, where
a positive test is defined as positive by PCR or by LFD followed by confirmatory PCR.  In
practice we further spatially aggregated Cornwall and the Isles of Scilly due to the
latter's small population size, and likewise City of London and City of Westminster for a
total of 380 discrete spatial units.

\begin{figure}
  \centering
  \includegraphics[width=0.55\textwidth]{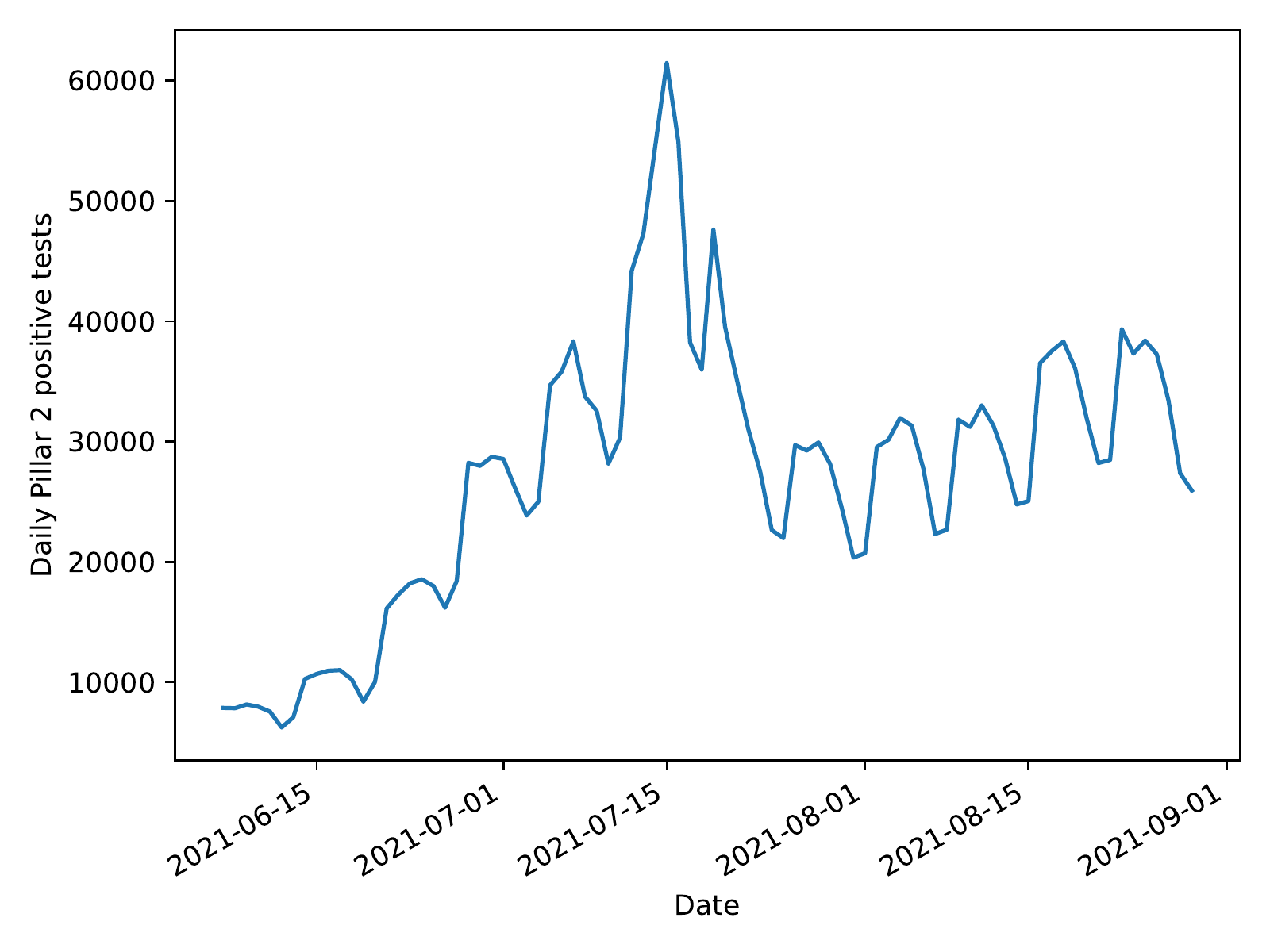}
  \hfill
  \includegraphics[width=0.4\textwidth]{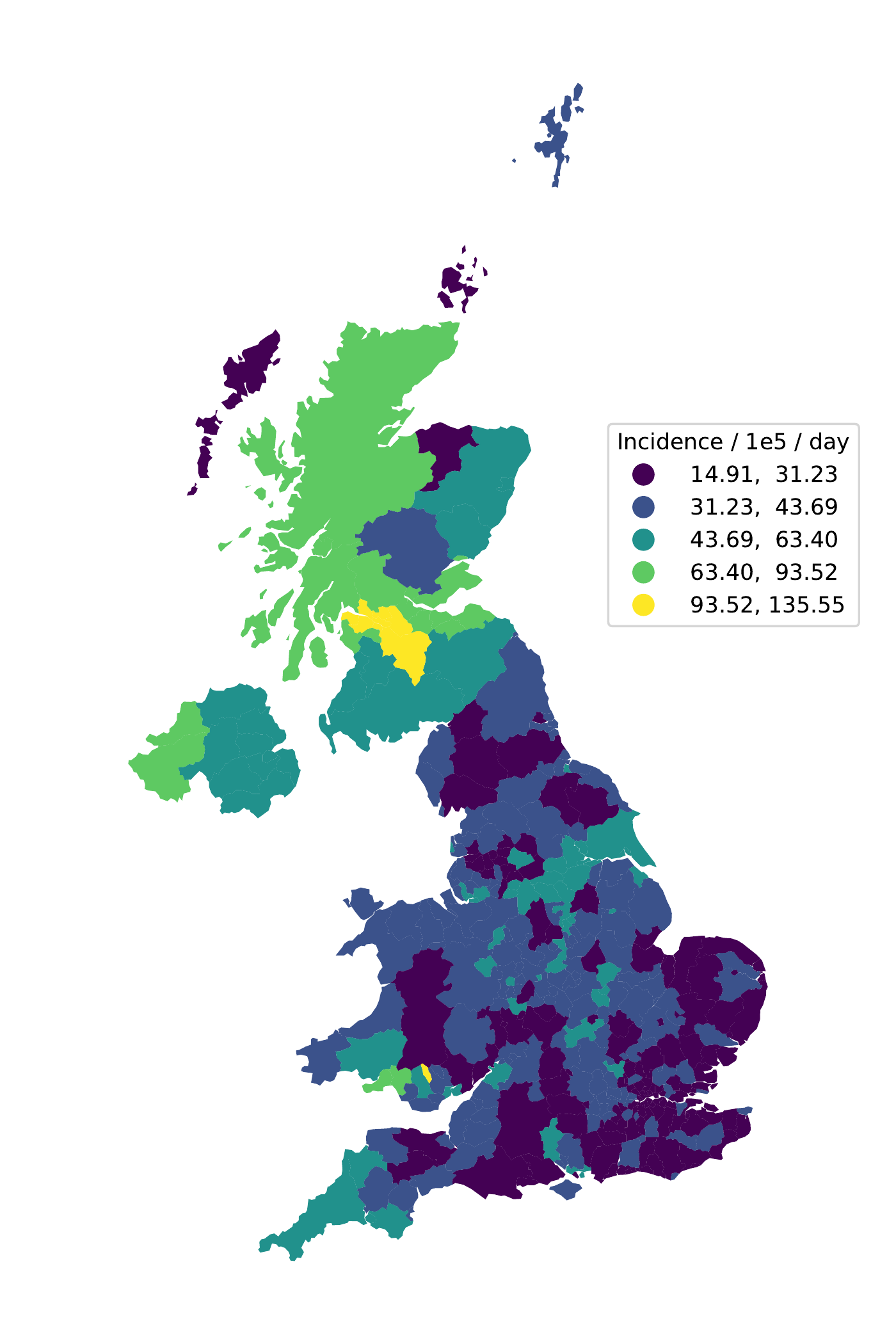}
  \caption{Cases of COVID-19 in the UK, as determined by daily numbers of Pillar 2 positive
  tests.  Left: COVID-19 daily case counts between 7th June and 29th August 2021 inclusive,
      showing marked weekly variation and long-term drift; Right: Spatial distribution of COVID-19 cases as of 29th August 2021, expressed as incidence per 100000 people per day.  Concentrations of high incidence are observed in Scotland, Northern Ireland, and Wales.}
  \label{fig:motivating-data}
\end{figure}

From May 2020 until March 2022, we ran our analysis daily on an 84 day sliding window.  In
practice, for an analysis on day $t$, we used data in the interval $[t-88, t-4)$,
discarding the latest 4 days' worth of data due to significant recording lag.  For the
purposes of exposition in this paper, however, we restrict our results to
the 84 day window between 7th June 2021 and 31th August 2021 inclusive.  This covered the
emergence of the SARS-CoV-2 Delta (B.1.617.2) strain, and was after social distancing
regulations had been relaxed. Figure \ref{fig:motivating-data} shows the daily case counts and overall incidence map for this period.

Connectivity between LADs was informed by freely available Census 2011
commuting volume data aggregated from Middle Super Output Area (MSOA) onto our 380
LAD-based spatial units.  These data provide
a matrix $G$ with $g_{ij}\; i,j=1,\dots,380$ being the number of journeys made from
“residence” $j$ to “workplace” $i$.  $G$ is non-symmetric, reflecting commuting behaviour rather
than the reciprocity of disease transmission.  We therefore calculated a symmetric matrix
$C = G + G^T$ of the daily number of journeys between each LAD assuming that commuters
return to their residence each day, and go from their residence to their workplace and
back at most once per day.  We also set $C_{ii} = 0, \; \mbox{for all}\; i$ as within-LAD
infection transmissibility is delegated to another part of our model (Section \ref{sec:model}).

Finally, we introduce LAD-level population estimates and LAD polygon area to inform how
transmission varies with population size.  In all datasets, two pairs of LADs (Cornwall
and Scilly, and City of Westminster and City of London) are merged to allow mapping of
MSOAs onto LADs resulting in data geolocated to 380 spatial units.

\section{Model}\label{sec:model}

In the following sections, we construct a model to represent the epidemic in terms of the
number of individuals testing positive on each day $t=0,\dots,T-1$ within spatial unit
$i=1,\dots,M$ defined by UK Local Authority Districts.  Since our motivation is to measure
quantities such as the degree of disease transmission between spatial units and reproduction
number, and predict the ongoing disease trajectory, we are interested in capturing
spatiotemporal dependence between disease prevalence and incidence.

Below, we first describe a linear state transition model that captures the natural history
of the COVID-19 disease progression per individual, accounting for segregation of the
population at risk into discrete spatial units.  We then describe how this model is set up
as a discrete time Markov process, establishing a notation allowing us to describe our
inference methodology in the next section.

\subsection{State transition model}\label{sec:state-transition-model}
Within each spatial unit, we model the COVID-19 epidemic by assuming that at any time
$t$ individuals exist in one of 4 mutually exclusive disease states: susceptible, exposed
(infected but not yet infectious), infectious, and removed which we denote by $\mathrm{S}$, $\mathrm{E}$, $\mathrm{I}$, and $\mathrm{R}$
respectively.  We assume that individuals progress sequentially between the disease
states, such that the allowed transitions are $\tx{S}{E}$, $\tx{E}{I}$, and $\tx{I}{R}$
as shown in Figure \ref{fig:seir-model}.  Each transition is associated with a hazard rate,
$\lambda^{\sidx{SE}}_{i}(t)$, $\lambda^{\sidx{EI}}_{i}(t)$, and $\lambda^{\sidx{IR}}_{i}(t)$ respectively giving the transition rate between states for a single individual, and
allowing us flexibility to specify how they evolve over space and time.

\subsubsection{Infection rate} The infection hazard rate $\lambda^{\sidx{SE}}_i(t)$ is
space- and time-dependent, parameterised as a product of a log-linear term describing the susceptibility
of an individual in $i$ to infection, and a function of infectious challenge assuming
homogeneous mixing of individuals within each spatial unit $i$ and mixing due to commuting
flows between $i$ and all other spatial units $j \ne i$.  We scale the hazard rate by the local
population size $n_i$ following the usual assumption of frequency-dependent disease
transmission.  Thus
\begin{equation}\label{eq:infection-rate}
\lambda^{\sidx{SE}}_i(t) = \frac{\exp (u_t + s_i)}{n_i} \left[x^{\sidx{I}}_{it} + \psi
\sum_{j \ne i} \frac{c_{ij}x^{\sidx{I}}_{jt}}{n_j} \right]
\end{equation}
where $u_t$ is a temporally-correlated random effect, $s_i$ a spatially-correlated random effect,
$x^{\sidx{I}}_{it}$ is the number of infectious individuals in $i$ at time $t$ (see below), and $c_{ij}$
represents the commuting flow to spatial unit $i$ from spatial unit $j$.  The coefficient $\psi$ is
assumed unknown and is to be estimated given the data.

Due to the considerable variability in the infection dynamics of the COVID-19 outbreak
we assume a random walk for the temporally-correlated random effect $\bm{u}$, such that
\begin{equation}\label{eq:u_t}
  u_t = \begin{cases}
          \alpha_0 & \mbox{if}\; t = 0 \\
          u_{t-1} + \alpha_t & \mbox{if}\; t>0
        \end{cases}
\end{equation}
with
\begin{eqnarray}
  \alpha_0 & \sim & \mbox{Normal}(0, 10) \\
  \alpha_{t} & \overset{\mbox{\tiny iid}}{\sim} & \mbox{Normal}(0, \sigma_u^2),\; t>0
\end{eqnarray}
and $\sigma_u = 0.005$ chosen heuristically.

Furthermore, we assume that the spatially-correlated random
effect follows a Conditional Autoregressive (CAR) process with joint
distribution
\begin{equation}\label{eq:s_i}
\bm{s} \sim \mbox{MVN}\left(\bm{0}, \Sigma\right).
\end{equation}
The covariance matrix $\Sigma$ is specified in terms of an adjacency matrix $W$ with
elements $(W)_{ij} = 1 \iff i\sim j$ and 0 otherwise, and a  diagonal matrix $D_w$ with
element $(D_w)_{ii} = W\cdot\mathbf{1}$ (i.e. the number of spatial units neighbouring
$i$) such that
\begin{equation}
  \Sigma = \sigma_s^2(D_w - \rho W)^{-1}
\end{equation}
with $\sigma_s$ an unknown parameter and correlation parameter $\rho = 0.25$ chosen heuristically to give a smooth posterior surface across LADs in an effort to improve reliability of model fitting by keeping the number of free parameters to a minimum.  We note that in a less operational context, a Besag-York-Molli\'e model might be preferred, though it would require the development of more complex and bespoke MCMC methodology to fit in the presence of latent event times.    

The following prior distributions are imposed on the remaining free parameters
\begin{eqnarray}\label{eq:priors}
  \psi & \sim & \mbox{Gamma}(0,100) \\
  \sigma_s & \sim & \mbox{HalfNormal}(0.1)
\end{eqnarray}

\subsubsection{Latent and infectious periods}
The latent period (sojourn in E) and
infectious period (sojourn in I) are parameterised through the transition rates
$\lambda^{\sidx{EI}}_i(t)$ and $\lambda^{\sidx{IR}}_i(t)$.   Evidence from the literature suggests a mean latent period of 5 - 7 days \citep{QuesEtAl21},  with a mean infectious period of 5 days starting 2 days before the onset of symptoms \citep{Tian20}.  In the context of our model, we therefore assume 
$$\lambda^{\sidx{EI}}_i(t) = 0.25$$
giving an \emph{effective} 4 day mean latent period irrespective of time and space.  For the infectious period, we assume
$$
\lambda^{\sidx{IR}}_i(t) = e^{\gamma_0 + \gamma_1 g_t}
$$
irrespective of space, where $\gamma_0 = \log 0.25$ is the (known) mean log hazard rate for removal ($\tx{I}{R}$) events, $\gamma_1$ the (unknown) log hazard ratio for being removed on a
weekday versus the weekend, and
$$
g_t = \begin{cases}-\frac{2}{7} & \mbox{if}\; t \in \{\mbox{mon},\dots,\mbox{fri}\} \\
                   \frac{5}{7} & \mbox{otherwise}
      \end{cases}
$$
to stabilise the inference.  We assume that \emph{a priori} $\gamma_1 \sim \mbox{Normal}(0, 100)$.

\subsection{Discrete-time Markov chain implementation}

In this section, we implement the SEIR state transition model described in
Section \ref{sec:state-transition-model} as a discrete-time stochastic process, more specifically
a discrete-time finite state-space Markov chain using the so-called ``chain binomial'' setup \citep[e.g.][]{becker1981}.

For timepoint $t$ and spatial unit $i$, let $x^{q}_{it}$ represent the number of individuals
resident in state $q \in \{\state{S}, \state{E}, \state{I}, \state{R}\}$, and
$y^{qr}_{it}$ for transition $[qr] \in \{\tx{S}{E},\tx{E}{I},\tx{I}{R}\}$ represent the number of transitions occurring between
each state.  Given an initial state $\bm{x}_0$, we evolve the epidemic by iterating
\begin{equation}
  y^{qr}_{it} \sim \mbox{Binomial}\left(x^{q}_{it}, p^{qr}_{i}(t)\right),\qquad [qr] \in
  \left\{\tx{S}{E},\tx{E}{I},\tx{I}{R}\right\}\label{eq:chain-binomial}
\end{equation} 
and
\begin{align}
  x^{\sidx{S}}_{i,t+1} & = x^{\sidx{S}}_{it} - y^{\sidx{SE}}_{it} \\
  x^{\sidx{E}}_{i,t+1} & = x^{\sidx{E}}_{it} + y^{\sidx{SE}}_{it} - y^{\sidx{EI}}_{it} \\
  x^{\sidx{I}}_{i,t+1} & = x^{\sidx{I}}_{it} + y^{\sidx{EI}}_{it} - y^{\sidx{IR}}_{it} \\
  x^{\sidx{R}}_{i,t+1} & = x^{\sidx{R}}_{it} + y^{\sidx{IR}}_{it} \label{eq:state-update}
\end{align}
where the transition probabilities $p^{qr}_{i}(t)=1 - e^{-\lambda^{qr}_i(t)\delta t}$ are assumed constant throughout each timestep of size $\delta t$.

Considering \eqref{eq:chain-binomial} as the data generating model, a realisation of the
Markov chain results in the $T\times M$ state matrices $\bm{x}^{q}$ and event matrices $\bm{y}^{qr}$
giving the state of the Markov chain at each timepoint and transition events responsible
for evolving the state.

\section{Inference}\label{sec:inference}

In order to fit our model to the spatial timeseries of positive COVID-19 tests across the UK, we make the assumption 
that the occurrence of a positive test is equivalent to observing a $\tx{I}{R}$ event.  
Furthermore, since the $\tx{S}{E}$ or $\tx{E}{I}$ events are censored, we adopt the conventional statistical notation 
$z^{\sidx{SE}}_{it} = y^{\sidx{SE}}_{it}$ and $z^{\sidx{EI}}_{it} = y^{\sidx{EI}}_{it}$ to denote the presence of censored data $\bm{z}$.

Given the data generating model given in Equations \eqref{eq:chain-binomial} and
\eqref{eq:state-update}, and conditional on unknown parameter vector $\bm{\theta} =
\{\psi, \sigma_s, \gamma_1, \bm{\alpha}, \bm{s} \}$ and initial conditions $\bm{x}_0$, the log
likelihood of observing transitions $\bm{y}$, censored transitions $\bm{z}$, and states
$\bm{x}_{1:T}$ given a set of initial conditions $\bm{x}_0$ and $\bm{\theta}$ is

\begin{align}\label{eq:likelihood} \ell(\bm{y}, \bm{z}, \bm{x} | \bm{x}_0, \bm{\theta}) =
\sum_{t=0}^{T-1} \sum_{i=1}^{M} [ & z^{\sidx{SE}}_{it} \log p^{\sidx{SE}}_{i}(t) +
(x^{\sidx{I}}_{it}-z^{\sidx{SE}}_{it})\log (1 - p^{\sidx{SE}}_{i}(t)) + \\ &
z^{\sidx{EI}}_{it} \log p^{\sidx{EI}}_{i}(t) + (x^{\sidx{E}}_{it}-z^{\sidx{EI}}_{it})\log (1
- p^{\sidx{EI}}_{i}(t)) + \\ & y^{\sidx{IR}}_{it} \log p^{\sidx{IR}}_{i}(t) + (x^{\sidx{I}}_{it} -
y^{\sidx{IR}}_{it})\log (1 - p^{\sidx{IR}}_{i}(t)) ] \\
\end{align}

We estimate the joint posterior of the parameters and missing data $\pi(\theta, \bm{z} |
\bm{y}, \bm{x}_0)$ by using a Metropolis-within-Gibbs MCMC scheme in which we draw alternately from
$\pi(\bm{\theta} | \bm{z}, \bm{y}, \bm{x}_0)$ by adaptive Hamiltonian Monte Carlo, and
$\pi(\bm{z} | \bm{\theta}, \bm{y}, \bm{x}_0)$ using a discrete-space Metropolis Hastings
step.  A high-level description of our approach is shown in Algorithm
\ref{alg:overall-mcmc}.

The adaptive Hamiltonian Monte Carlo algorithm for drawing from $\pi(\bm{\theta} | \bm{z},
\bm{y}, \bm{x}_0)$ is well-described elsewhere \citep[see for example][]{STAN}, and
therefore in the following sections we focus on describing the methodology for drawing
from $\pi(\bm{z} | \bm{\theta}, \bm{y}, \bm{x}_0)$ and $\pi(\bm{x}_0, \bm{z} | \bm{\theta}, \bm{y})$.

\begin{algorithm}
\DontPrintSemicolon
Initialise $\bm{\theta}^{(0)}, \bm{z}^{(0)}$ \;

\For{$k$ in $1,\dots,K$}{
Update $\bm{\theta}$ by HMC \;
\For{$l$ in $1,\dots,L$}{
    Update partially-censored events in $\bm{z}^{\sidx{SE}}$ \label{step:moveSE}\;
    Update partially-censored events in $\bm{z}^{\sidx{EI}}$ \label{step:moveEI}\;
    Update occult events in  $\bm{z}^{\sidx{SE}}$ \label{step:occultSE}\;
    Update occult events in $\bm{z}^{\sidx{EI}}$ \label{step:occultEI}\;
    Update initial conditions $\bm{x}_0$ for events $\bm{z}^{\sidx{SE}}$ \label{step:initSE}\;
    Update initial conditions $\bm{x}_0$ for events $\bm{z}^{\sidx{EI}}$ \label{step:initEI}\;
    }
}
\caption{Overall Metropolis-within-Gibbs MCMC algorithm for sampling from the joint posterior.}\label{alg:overall-mcmc}
\end{algorithm}

\subsection{Drawing samples from $\pi(\bm{z} | \bm{\theta}, \bm{y}, \bm{x}_0)$}\label{sec:censored-data}

In this section we describe the discrete-space Metropolis-Hastings algorithms used to draw
samples from the conditional posterior of the censored events and initial conditions given state transition parameters and data, $\pi(\bm{z}, \bm{x}_0 | \bm{\theta}, \bm{y})$.
The approach follows the principles for data-augmentation in individual level models
\citep[see for example][]{JewEtAl09c}, which involves two separate Metropolis-Hastings
kernels to explore the space of partially-censored and fully-censored events respectively.  Unlike in the continuous-time setting, for our discrete model we also introduce a third Metropolis-Hastings kernel which explores the initial conditions space.

In the context of our model, the challenge to successful updating of censored event times
is to create a proposal mechanism that respects the constraint that $x^{q}_{it} \geq 0$
for all $q \in \{\mathrm{S}, \mathrm{E}, \mathrm{I}, \mathrm{R}\}$ and $t=0,\dots,T-1$ and
$i=1,\dots,M$.

\subsubsection{Updating partially-censored events}\label{sec:update-partially-observed-events}

The proposed Metropolis Hastings kernel operates on a $M\times T$ censored event matrix
$\bm{z}^{qr}, \; [qr] \in \{\tx{S}{E}, \tx{E}{I}\}$.  The algorithm proceeds by
proposing to move a number of events $w$ from $z^{qr}_{it}$ to $z^{qr}_{it+d}$ where we draw
\begin{align}
    i,t & \sim \mbox{Discrete}\left(i,t:z^{qr}_{it} > 0\right) \label{eq:move-draw-it} \\
    d & \sim \mbox{Discrete}\left(0 \vee t - d_{\mathrm{max}},\dots,-1, 1,\dots, T \wedge t + d_{\mathrm{max}}\right) \label{eq:move-draw-delta} \\
    w & \sim \mbox{Discrete}\left(1,\dots, B_1(z^{qr}_{it})\right) \label{eq:move-draw-b}
\end{align}
where $d_{max} > 0$ is a tuning constant, and with bounding function
\begin{equation}
    B_1(z^{qr}_{it}) = z^{qr}_{it} \wedge w_{\max} \wedge \begin{cases} \min (x^{r}_{it}, \dots, x^{r}_{i,t+d-1}) & \mbox{if} \; d > 0 \\
                                                 \min(x^{q}_{i,t-d}, \dots, x^{q}_{t-1}) & \mbox{if} \; d < 0
                                   \end{cases}
\end{equation}
where $w_{\max} > 0$ is also a tuning constant.

We then propose
\begin{align}
    z^{qr\star}_{it} & = z^{qr}_{it} - w \\
    z^{qr\star}_{t+\delta,i} & = z^{qr}_{t+\delta} + w
\end{align}
and accept the proposal with probability
\begin{equation}
    \alpha(\bm{z}^{qr}, \bm{z}^{qr\star}) = \frac{\pi(\bm{z}^{qr\star} | \bm{\theta}, \bm{y}, \bm{x}_0)}
                                       {\pi(\bm{z}^{qr} | \bm{\theta}, \bm{y}, \bm{x}_0)}
                                  \cdot
                                  \frac{\nnz{\bm{z}^{qr\star}}B_1(z^{qr\star}_{it}; \bm{x}^\star)}{\nnz{\bm{z}^{qr}}B_1(z^{qr}_{it}; \bm{x})} 
                                  \wedge 1
\end{equation}
where $\nnz{\bm{z}^{qr}}$ denotes the number of non-zero elements in $\bm{z}^{qr}$.

\subsubsection{Updating occult events}

To explore the space of occult events, we again operate on $M\times T$ censored event
matrices $\bm{z}^{qr},\; \tx{q}{r} \in \{\tx{S}{E}, \tx{E}{I}\}$, proposing to add or delete a
number of events to randomly chosen elements.  In our model, occult events are
overwhelmingly likely to occur close to the end of the analysis time-window, and so we
limit our choice of elements to those in the last 21 days of the timeseries.

To add events, we choose an element $z^{qr}_{it}$ to update via
\begin{align}
    i & \sim \mbox{Discrete}(1,\dots,M) \\
    t & \sim \mbox{Discrete}(T-21,\dots, T)
\end{align}
and propose to add $v$ events by
\begin{equation}
    v \sim \mbox{Discrete}(1,\dots, B_2(\bm{z}^{qr}_{it})),
\end{equation}
where bounding function
\begin{equation}
    B_2(z^{qr}_{it}) = v_{\max} \wedge \min (x^{q}_{it}, \dots, x^{q}_{iT})
\end{equation}
with $v_{\max}$ a tuning constant.

We then update $z^{qr\star}_{it} = z^{qr}_{it} + v$, and accept the proposal with probability
\begin{equation}
    \alpha(\bm{z}^{qr}, \bm{z}^{qr\star}) = \frac{\pi(\bm{z}^{qr\star} | \bm{\theta}, \bm{y}, \bm{x}_0)}
                                       {\pi(\bm{z}^{qr} | \bm{\theta}, \bm{y}, \bm{x}_0)}
                                  \cdot
                                  \frac{21MB_2(\bm{z}^{qr})}{\nnz{\bm{z}^{qr\star}}B_3(\bm{z}^{qr\star})} \wedge 1
\end{equation}
with $B_3(\bm{z}^{qr\star})$ defined below.

To delete events, we restrict our choice of $i$ and $t$ to positive elements of $\bm{z}^{qr}$ as in Equation \eqref{eq:move-draw-it}, and propose
\begin{align}
    i, t & \sim \mbox{Discrete}(i, t: z^{qr}_{it}>0, t \geq T-21) \\
    v & \sim \mbox{Discrete}(1, \dots, B_3(z^{qr}_{it}; \bm{x}))
\end{align}
where bounding function
\begin{equation}
    B_3(z^{qr}_{it}) = z^{qr}_{it} \wedge v_{\max} \wedge \min (x^{r}_{it}, \dots, x^{r}_{iT}).
\end{equation}

We then accept the proposal $z^{qr\star}_{it} = z^{qr}_{it} - v$ with probability
\begin{equation}
    \alpha(z^{qr}, z^{qr\star}) = \frac{\pi(\bm{z}^{qr\star} | \bm{\theta}, \bm{y}, \bm{x}_0)}
                                       {\pi(\bm{z}^{qr} | \bm{\theta}, \bm{y}, \bm{x}_0)}
                                      \cdot
                                  \frac{\nnz{\bm{z}^{qr}}B_3(\bm{z}^{qr})}{21MB_2(\bm{z}^{qr\star})}
                                  \wedge 1.
\end{equation}

\subsubsection{Updating initial conditions}
In the third data-augmentation kernel, we update the initial conditions matrix $\bm{x}_0$.
Given that we assume a closed population, such that $\bm{x}^{\sidx{S}} + \bm{x}^{\sidx{E}} +
\bm{x}^{\sidx{I}} + \bm{x}^{\sidx{R}} = \bm{N}$ the size vector of the population, we have two options for
updating $\bm{x}_0$.  The first would be to swap individuals between pairs of adjacent
epidemiological states.  However, since $\bm{x}_0$ is \emph{a posteriori} conditional on the
 events $\bm{z}$, the small conditional variance $Var(\bm{x}_0 | \bm{z}, \bm{y}, \bm{\theta})$ results in
slow exploration of the censored data space.  Instead, we employ a \emph{joint} update of
$\bm{x}_0$ and $\bm{z}$ in which we consider a set of left-censored events occurring prior to
timepoint $t=0$ that have given rise to $\bm{x}_0$.

For a given transition $\tx{q}{r}$, we begin as before by choosing an element $z^{qr}_{it}$ to update 
\begin{align*}
i & \sim \mbox{Discrete}(1,\dots,M) \\
t & \sim \mbox{Discrete}(0,\dots,6).
\end{align*}
noting that we restrict the choice of $t$ to a window $[0, 7)$ at the beginning of
the epidemic -- this helps to reduce the possible dimensionality of the discrete random
walk, since it is unlikely that left-censored events could be moved to later timepoints
without large changes in the posterior leading to a rejected move.

We now choose to either move events forwards ($d=1$) or backwards ($d=-1$) in time with equal probability such that
$$
d \sim \mbox{Discrete}(-1, 1)
$$
and a number of events to move such that
\begin{equation}\label{eq:init-conditions-h}
h \sim \mbox{Discrete}(1,\dots,B_4(\bm{x}_0))
\end{equation}
with
$$
B_4(\bm{x}_0) =  h_{\max} \wedge
\begin{cases}
  z^{qr}_{it} \wedge \min(x^q_{i1},\dots,x^q_{it}) & \mbox{if}\; d = -1 \\
  \min(x^r_{i0}, \dots,x^r_{it-1}) & \mbox{if}\; d = 1)
\end{cases}
$$
and where $h_{\max} > 1$ is a tuning constant.
We then let
\begin{align*}
x^{q\star}_{i0} = x^q_{i0} + d h \\
x^{r\star}_{i0} = x^r_{i0} - d h \\
z^{qr\star}_{it} =  z^{qr}_{it} + d h
\end{align*}
and accept the proposal with
$$
\alpha(\bm{x}_0, \bm{z}, \bm{x}_0^\star, \bm{z}^\star) = 
  \frac{\pi(\bm{x}_0^\star, \bm{z}^\star | \bm{y}, \bm{\theta})}
  {\pi(\bm{x}_0, \bm{z} | \bm{y}, \bm{\theta})} 
  \cdot
  \frac{B_4(\bm{x}_0^\star)}{B_4(\bm{x}_0)}\wedge 1.
$$
We note that if $B_4 = 0$, then $f_k(h) = 0\; \forall\; h$ in Equation \ref{eq:init-conditions-h} and the move will be rejected.

\subsection{Definition of $R_t$}

We define an approximate stratified temporal reproduction number $R_{jt}$ as the expected
number of further individuals that an individual in stratum $j$ will go on to infect given
the state at time $t$.  We define this by first considering the pairwise force of
infection defined by $\bm{K}$ exerted by an individual in $j$ on a susceptible individual
in $i$ such that
\begin{equation} \bm{R}_{t} \approx \frac{1 - \exp \left(-\left( \bm{X}^{\sidx{S}}_{\cdot
          t} \right)^T \cdot \bm{K} \right)}{1 - \exp(-\gamma_0)}
  \label{eq:Rjt}
\end{equation} We remark that this is an approximation since both
$\bm{X}^{\sidx{S}}_{\cdot t}$ (and therefore $\lambda^{\sidx{SE}}(t)$) is assumed constant
over the course of an individual's infectious period.

\subsection{Software implementation}
The model and sampler code were implemented in Python 3.8 using TensorFlow and TensorFlow Probability computational and probabilistic programming libraries for GPU acceleration \citep{tensorflow2015-whitepaper, tensorflow-probability}.  The Python code implementing this analysis is freely available under the MIT license at \url{https://gitlab.com/chicas-covid19/covid19uk}, with the Version 1.0 snapshot available at \url{https://doi.org/10.5281/zenodo.7715657}.

\section{Sampler optimisation}\label{sec:optimisation}

Before addressing the real-world application, we demonstrate that the discrete-space
samplers described above are optimised at the conventional accept/reject ratio of 0.23.
Here we provide results for tuning $m$, the number of metapopulations and $w$, the number
of events for the partially-observed event time moves described in Section
\ref{sec:update-partially-observed-events}, as applied to the $\tx{S}{E}$ and $\tx{E}{I}$
transitions respectively.

We first define the mean squared jumping distance to be
$$
\mbox{MSJD} = \frac{1}{K} \sum_{k=1}^{K} \sum_{i,t,s} \left(x^{s(k)}_{it} - x^{s(k-1)}_{it}\right)^2
.$$

We then run the partially-observed event time sampler separately for the $\tx{S}{E}$ and $\tx{E}{I}$ transitions respectively, considering all other censored data and parameters fixed at values taken from a randomly-chosen iteration of the converged MCMC chain described in Section \ref{sec:application}.  For each transition, we run the sampler for $(m, x_{max}) \in \{m: 1,\dots, 10\} \times
\{x_{max}: 1, \dots, 100 \}$ with $m$ and $x_{max}$ as defined in Section \ref{sec:inference}. 

\begin{figure}
  \centering
  \includegraphics[width=0.45\textwidth]{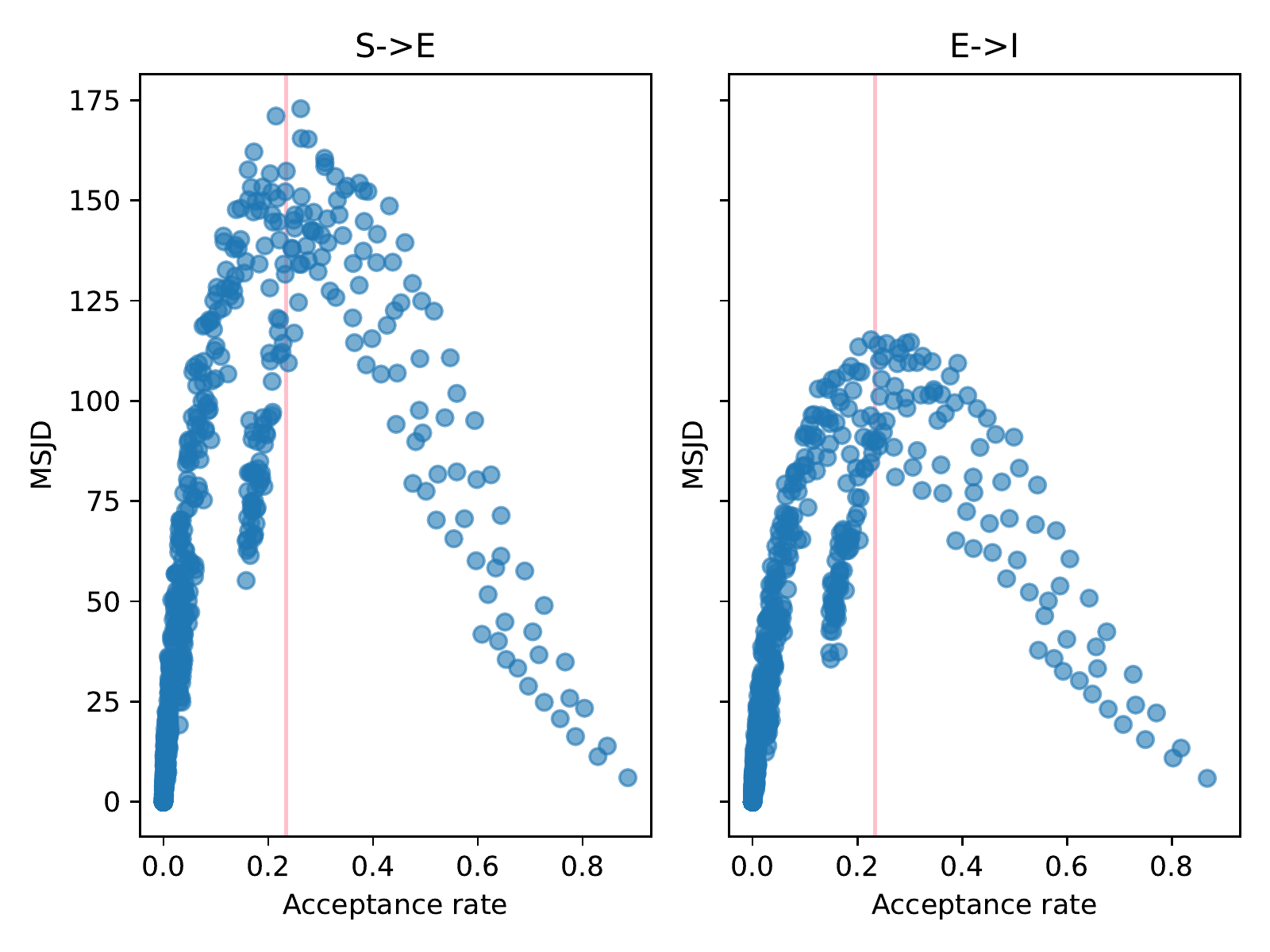}
  \caption{Mean squared jumping distance versus acceptance probability for the $\tx{S}{E}$
    (left) and $\tx{E}{I}$ (right) optimisation studies respectively (Section \ref{sec:optimisation}).  Blue dots represent each combination of $m$ and $x_{max}$, and red vertical line indicates theoretically optimal acceptance rate.}
  \label{fig:msjd_acceptance}
\end{figure}

For each transition and $(m$, $x_{max})$ tuple, we plot the MSJD against the acceptance
ratio in Figure \ref{fig:msjd_acceptance}.  This confirms that the MSJD is
maximised at an acceptance ratio of approximately 0.234 irrespective of the transition.
However, the way in which $(m$, $x_{max})$ affects the MSJD differs depending on the
transition selected.  Moreover, we observe that the total number of events moved across all selected metapopulations is conserved, such that as long as $mx_{max} \approx 30$ we achieve the maximum MSJD.

\begin{figure}
  \centering
  \includegraphics[width=0.7\textwidth]{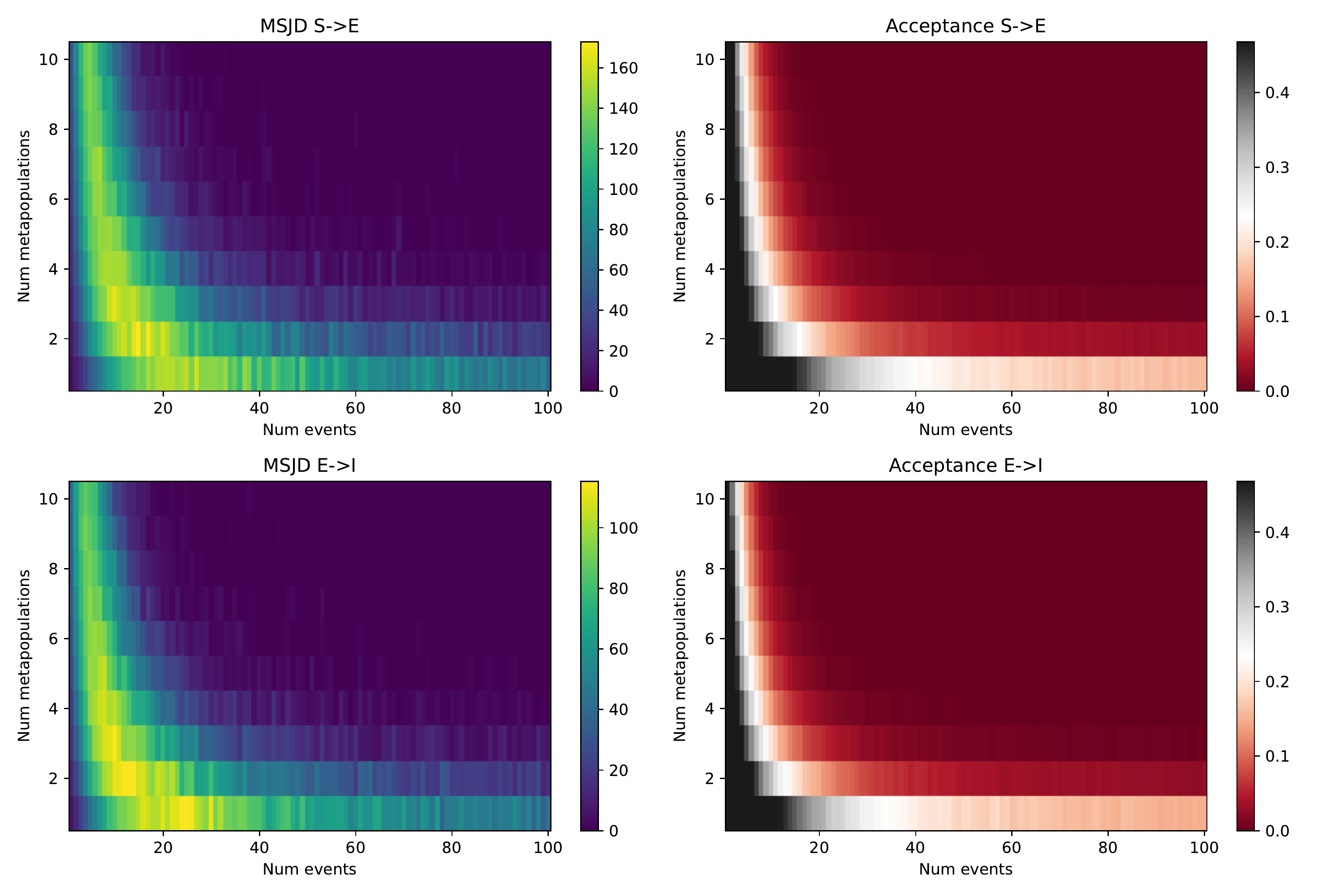}
  \caption{Mean squared jumping distance (left) and Metropolis Hastings acceptance
    probability (right)
    for the \emph{move} algorithm for both the $\tx{S}{E}$ (top) and $\tx{E}{I}$ (bottom)
    transitions.}
  \label{fig:m_n_optimisation.pdf}
\end{figure}

\section{Application to the UK COVID-19 epidemic}\label{sec:application}

In this section we apply the model described in Section \ref{sec:model} to the UK COVID-19 spatial case timeseries shown in Section \ref{sec:motivating-dataset}.  The MCMC algorithms were run for $k=40000$ iterations, with $l=380$ censored event updates per iteration (Algorithm \ref{alg:overall-mcmc}) with tuning constants as in Table \ref{tab:tuning-constants}.  Three independent MCMC algorithms were run, initialising each Markov chain using a random draw from the prior distribution, and discarding the first 10000 iterations for any calculated quantities ($R_t$, riskmaps, and predictive distributions).

\begin{table}
  \caption{\label{tab:tuning-constants}Tuning constants used for MCMC algorithm to fit the spatial stochastic meta-population model to COVID-19 case timeseries data.}
  \centering
  
  \begin{tabular}{lllr}
  \hline
  \textbf{Transition} & \textbf{Update} & \textbf{Tuning constant} & \textbf{Value} \\
  \hline
    \multirow[t]{4}{*}{$\tx{S}{E}$} & \multirow[t]{2}{*}{Partially-censored}
     & $d_{max}$ & 84 \\
     & & $w_{max}$ & 22 \\
     & Fully-censored & $v_{max}$ & 20 \\
     & Initial conditions & $h_{max}$ & 19 \\
  \hline
    \multirow[t]{4}{*}{$\tx{E}{I}$} & \multirow[t]{2}{*}{Partially-censored}
     & $d_{max}$ & 84 \\
     & & $w_{max}$ & 18 \\
     & Fully-censored & $v_{max}$ & 20 \\
     & Initial conditions & $h_{max}$ & 17 \\
     \hline    
  \end{tabular}
\end{table}

\begin{figure}  \centering
  \includegraphics[width=0.65\textwidth]{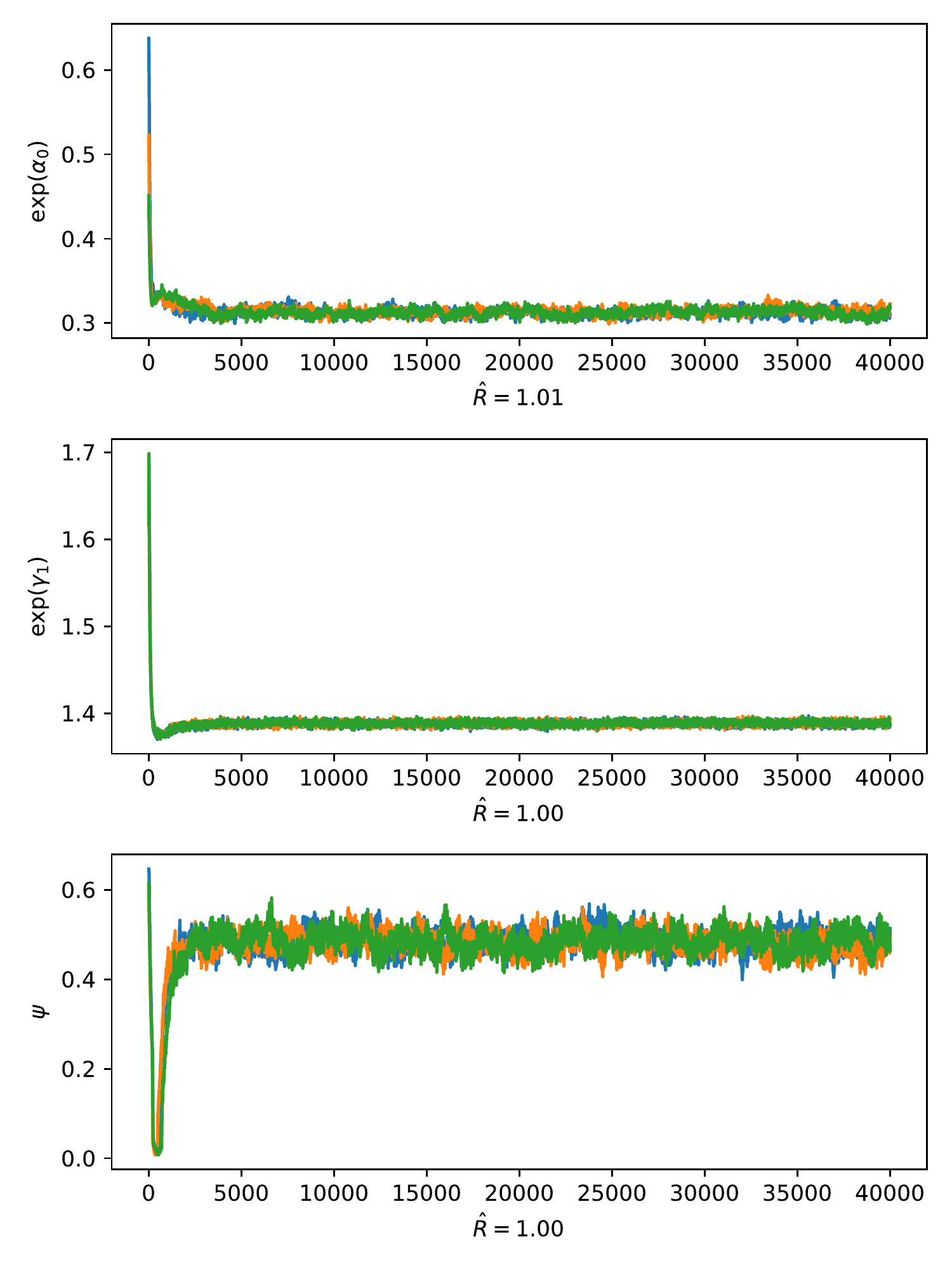}
  \caption{Traceplots of uni-dimensional parameters $\exp(\alpha_0)$ (i.e. $\alpha_0$ expressed as absolute infection risk per individual per day), $\exp(\gamma_1)$ (i.e. $\gamma_1$ expressed as
  relative risk), and $\psi$ (infection rate per commute visit).}
  \label{fig:plot_spatial_samples_traces_all}
\end{figure}

Traceplots of the scalar quantities $\exp(\alpha_0)$, $\exp(\gamma_1)$, $\psi$ are shown in Figure \ref{fig:plot_spatial_samples_traces_all}, superimposing the three independent chains and calculating the Brookes-Gelman-Rubin statistics ($\hat{R}$, not to be confused with the time-varying reproduction number $R_t$) for each parameter  \citep{BrGel98}.  The algorithm exhibits satisfactory convergence for all three chains, after a conservative 10000 iteration burn-in which is removed to compute the following now-casting and predictive results. 

\begin{figure}  \centering
  \includegraphics[width=\textwidth]{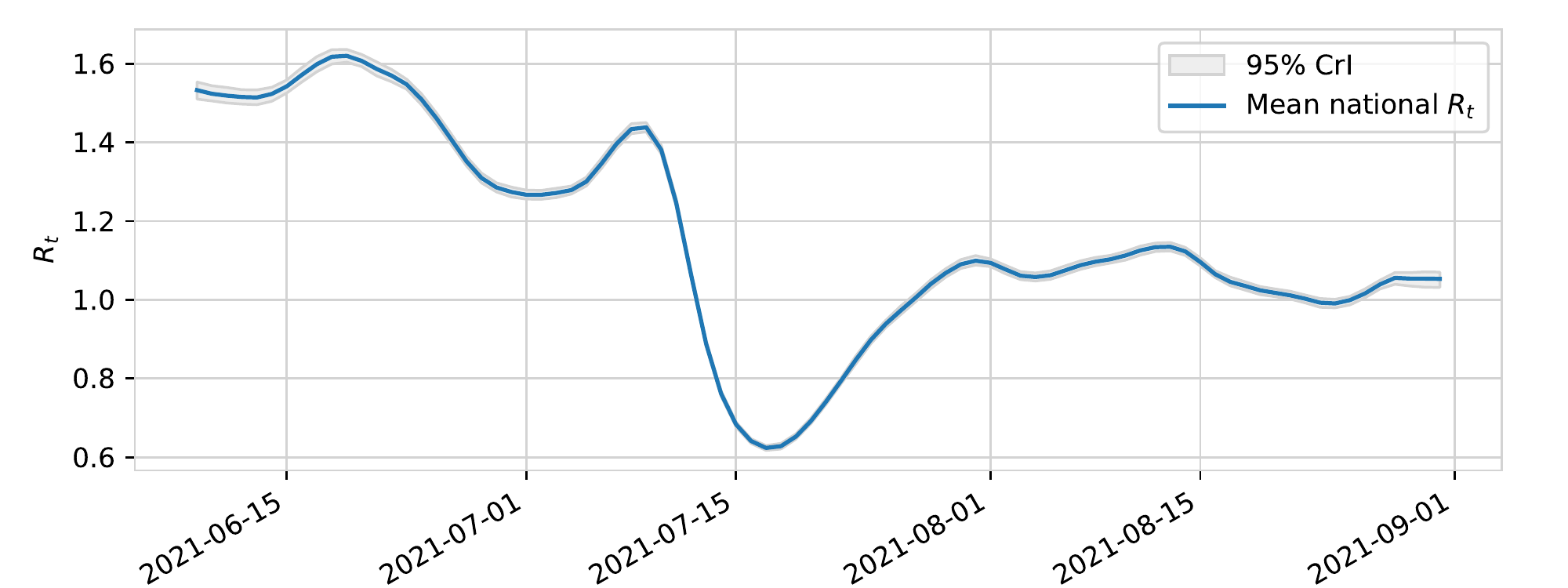}
  \caption{Posterior distribution of $R_t$ over the analysis period.}
  \label{fig:Rt}
\end{figure}

The temporal trend in the global time-varying reproduction number is shown in Figure \ref{fig:Rt}, demonstrating marked variation through time either side of $R_t=1$ as expected given the case timeseries trajectory in Figure \ref{fig:motivating-data}.  The credible intervals around $R_t$ are narrow, reflecting the choice of case observation model and the large number of daily cases occurring in the UK as a whole, with variation due to spatial connectivity and risk accounted for by the model. 

\begin{figure} \centering
  \includegraphics[width=0.45\textwidth]{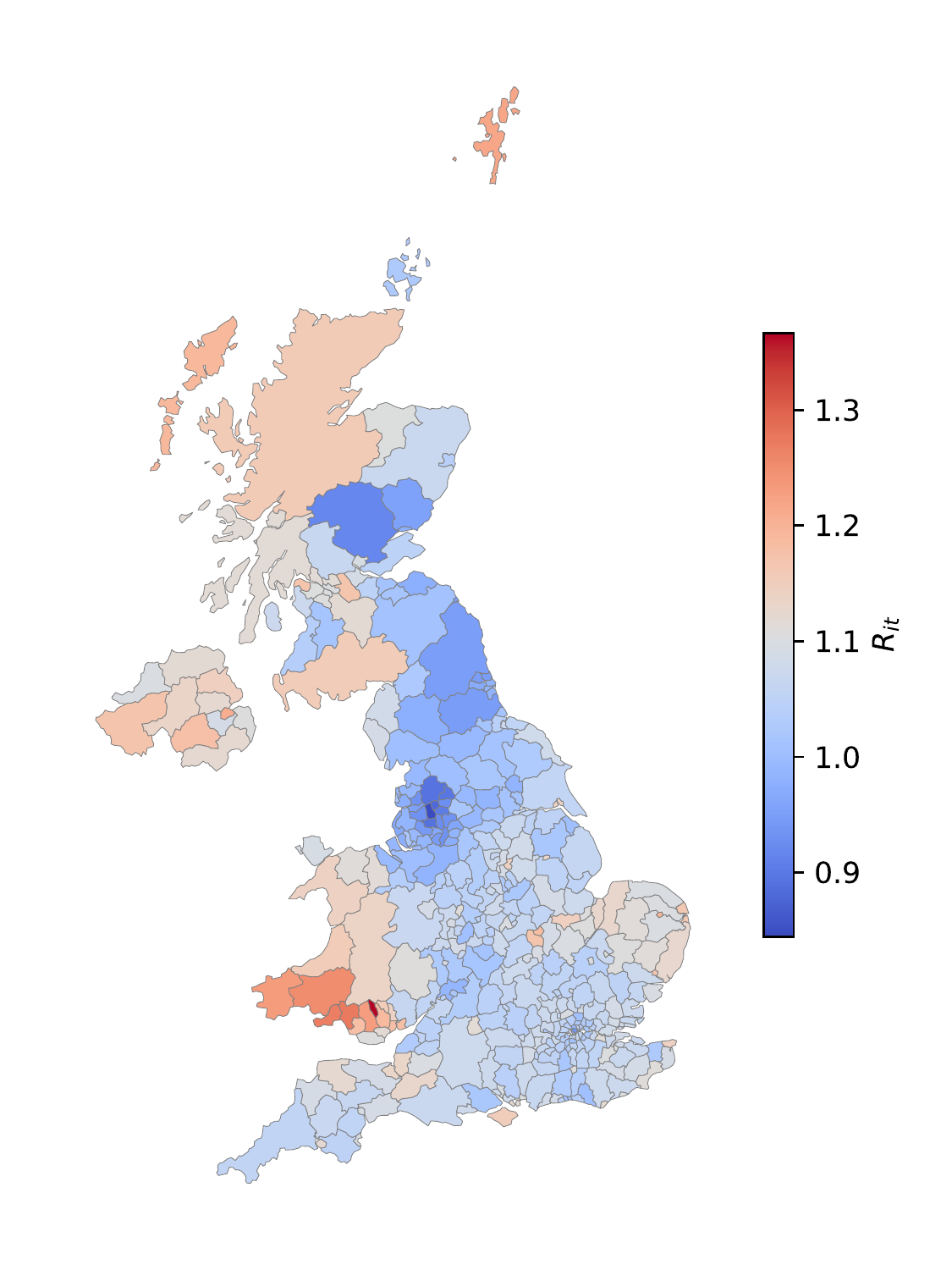}
  \includegraphics[width=0.45\textwidth]{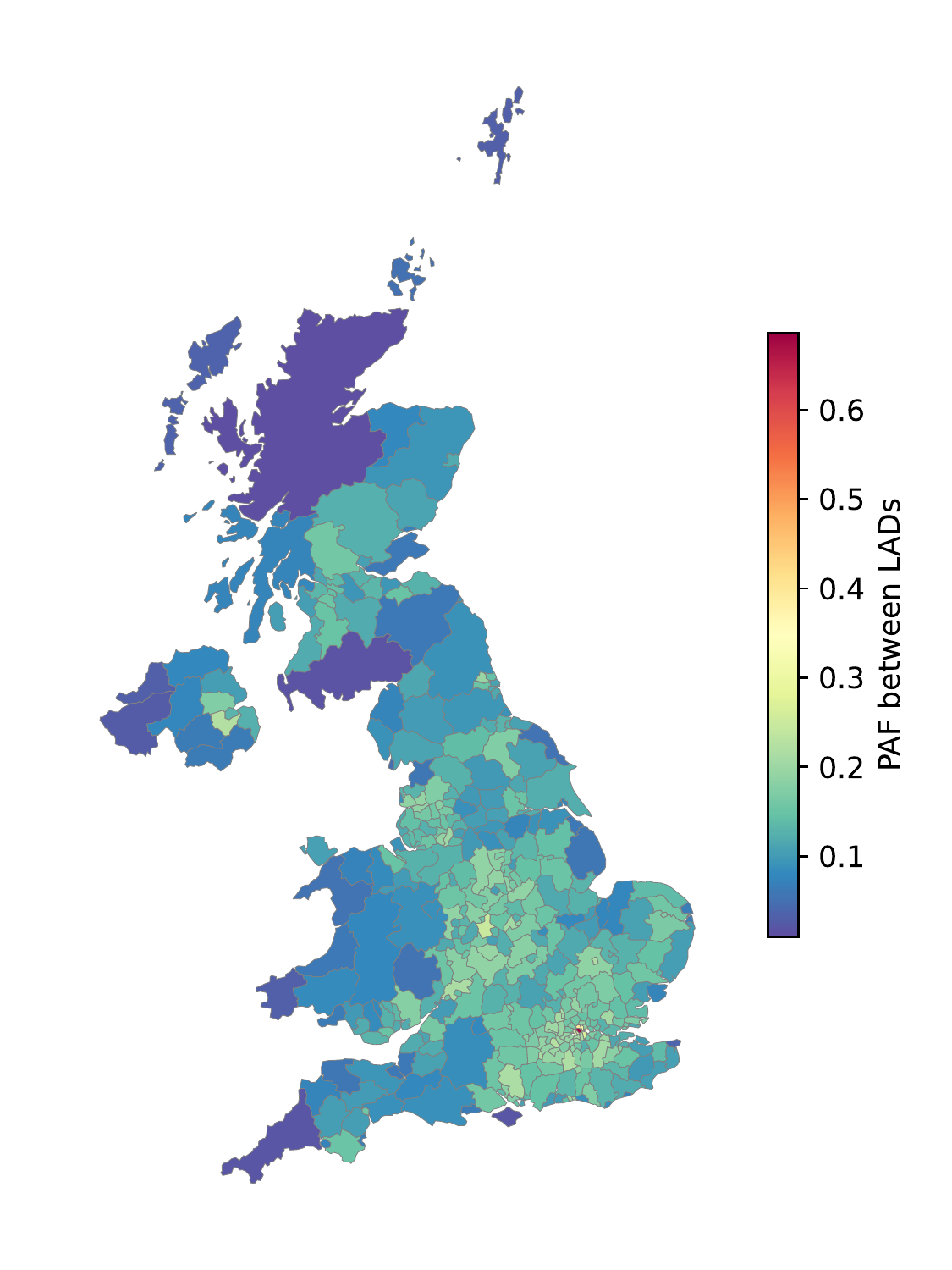}
  \caption{Posterior mean spatial reproduction number $R_{it}$ (left) and Population Attributable Fraction of the infection hazard due to between-LAD
    mobility (right) as of 31st August 2021.}
  \label{fig:spatial_posterior_Rt_map}
\end{figure}

At the LAD-level, Figure \ref{fig:spatial_posterior_Rt_map} (left) shows the mean posterior local reproduction number $R_{it}$ as of the 31st August 2021.  Considerable spatial heterogeneity is seen, driven only in part by the spatial distribution of cases on the same date (Figure \ref{fig:motivating-data}).  This reflects the dynamics of the spatial epidemic over the whole time-window, with areas such as East Anglia exhibiting higher $R_{it}$ values than might be concluded simply by looking at the most recent case data.

A convenient way to study the importance of spatial transmission is to plot the attributable fraction of the total infection risk on an individual in LAD $i$ which is due to between-LAD transmission, defined as
\begin{equation}\label{eq:af}
AF_{it} = \frac{\phi \sum_{j \ne i} \frac{c_ij x^{\sidx{I}}_{jt}}{n_j}}{x^{\sidx{I}}_{it} + \phi \sum_{j \ne i} \frac{c_ij x^{\sidx{I}}_{jt}}{n_j}}.
\end{equation}
For a particular LAD of interest, the importance of spatial transmission therefore varies with both connectivity and within-LAD disease prevalence.  For the most recent timepoint in our case timeseries, we plot the posterior mean AF per LAD.  Highly connected LADs, such as those comprising or close to the major cities, typically have a higher AF than more rural LADs, driven by overall population mobility.

\begin{figure}  \centering
  \includegraphics[width=0.45\textwidth]{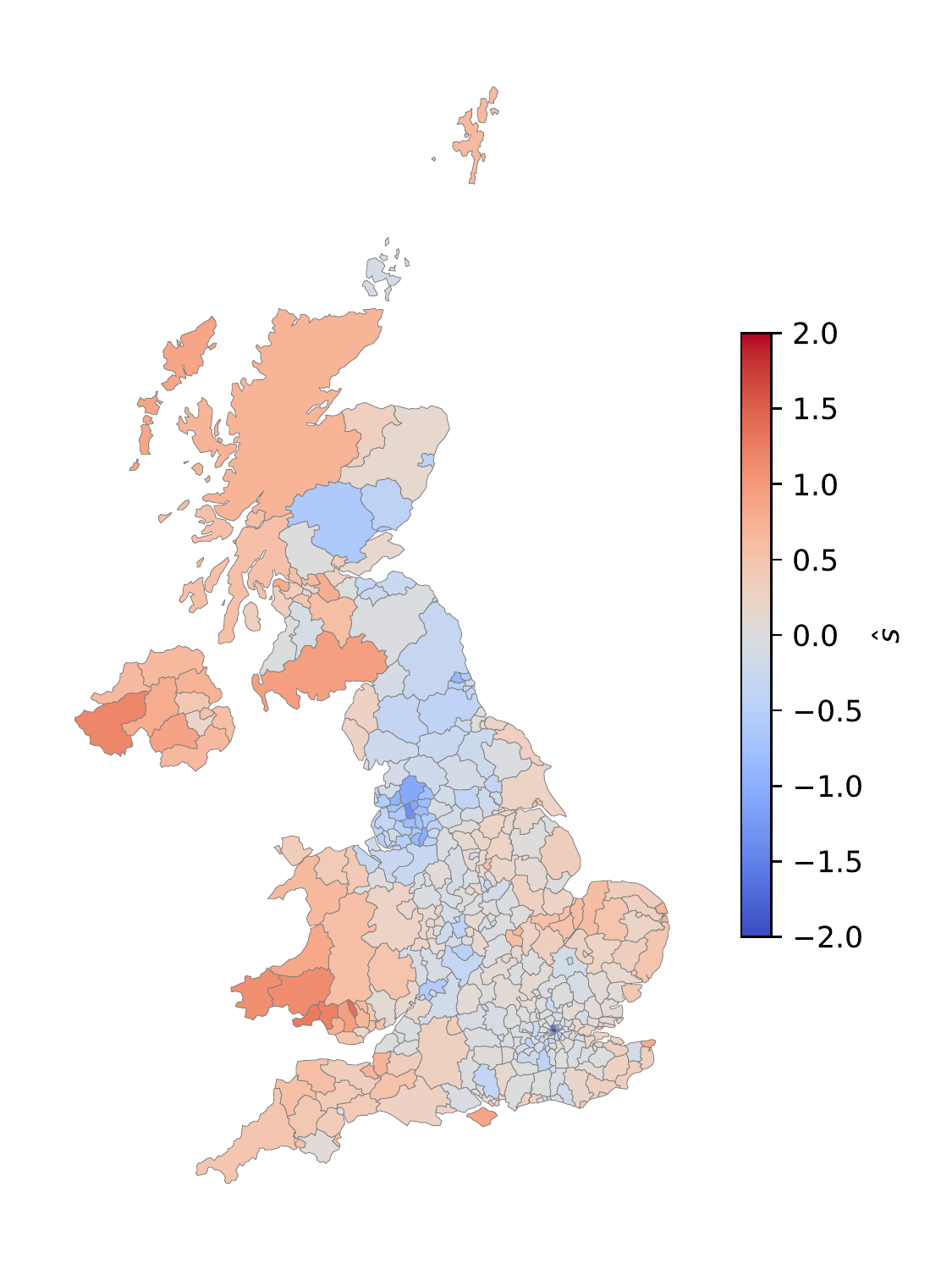}
  \includegraphics[width=0.45\textwidth]{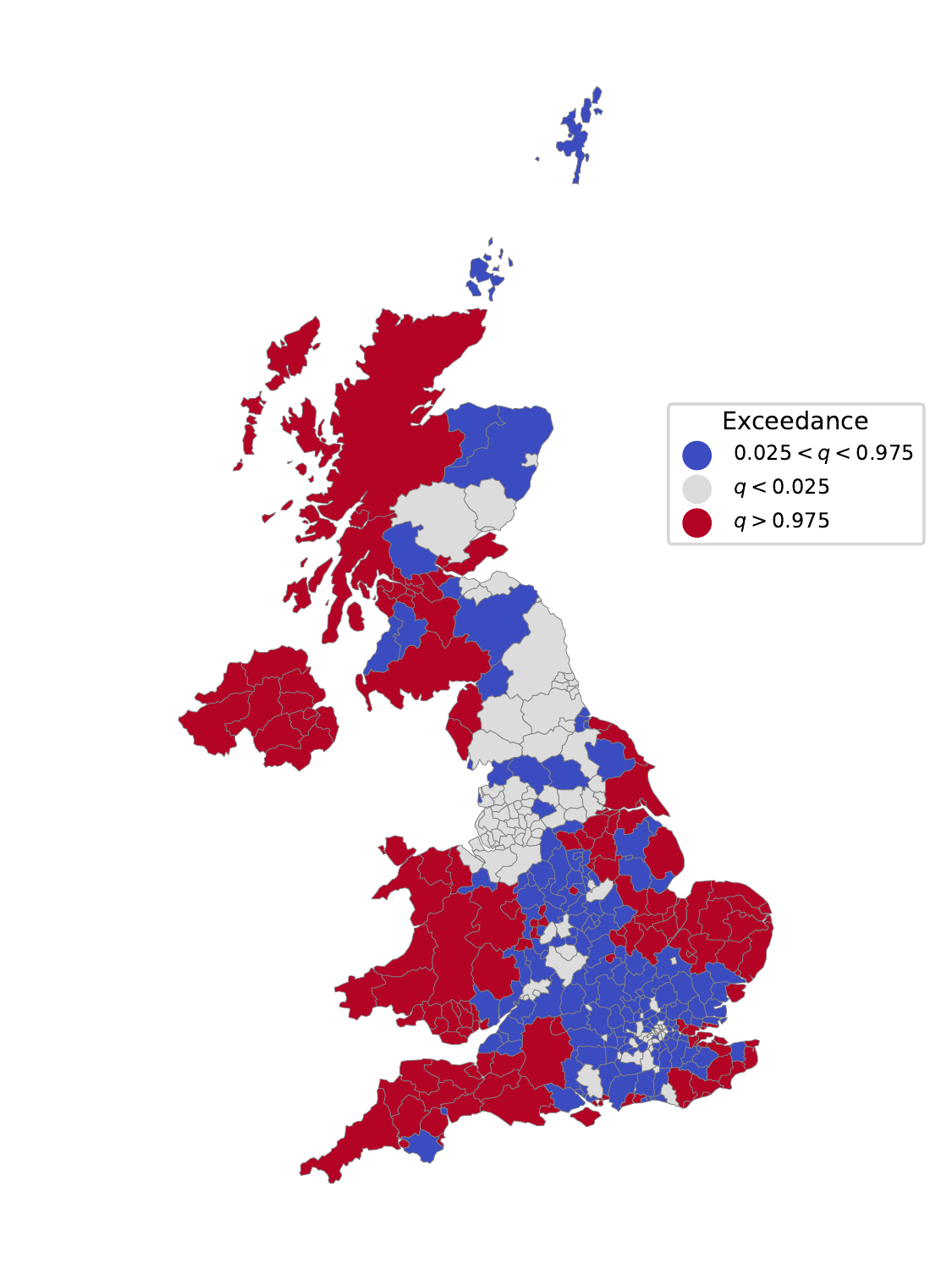}
  \caption{Posterior mean of $\bm{s}$ (left), and exceedance probability $q = Pr(s>0)$ (right) for each Local
  Authority District.}
  \label{fig:spatial-posterior}
\end{figure}

Whilst our model captures the national-level baseline transmission rate and inter-LAD connectivity, it also allows local variation in baseline transmission rate through our spatially-correlated term $\bm{s}$.  The posterior mean of $\bm{s}$ and exceedance probability $Pr(s > 0)$  is plotted geographically in Figure \ref{fig:spatial-posterior}.  In general, these results indicate a trend towards higher baseline transmission in less densely populated, rural LADs compared to more populated, urban LADs.

\begin{figure} \centering
  \includegraphics[width=\textwidth]{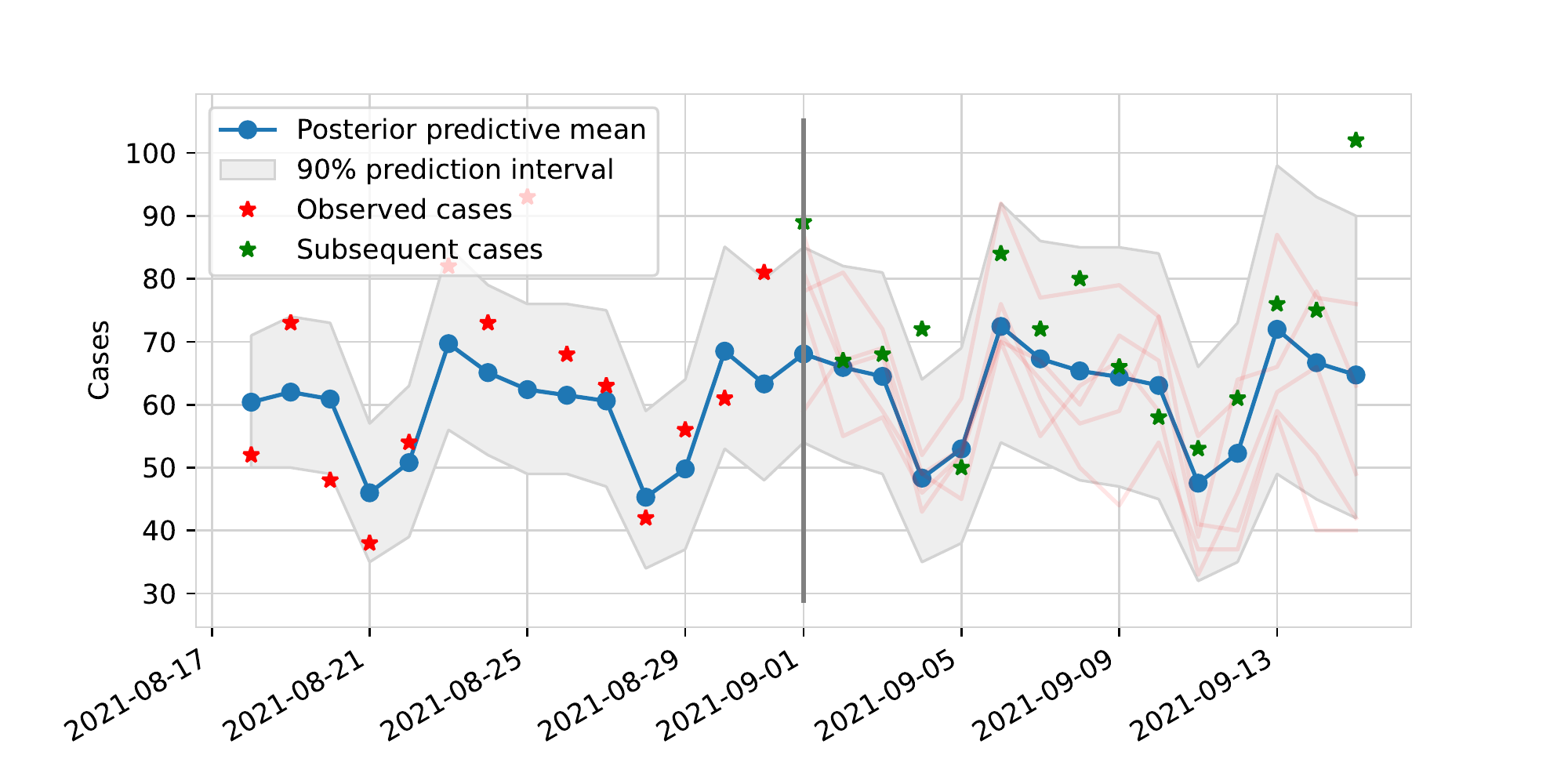}
  \caption{\label{fig:predictive-plot}Smoothing distribution (left of vertical line) and predictive distribution
    (right of vertical line) of number of $\tx{I}{R}$ events.  The smoothing distribution
    is compared against the observed case data for the Lancaster LAD for the last 2 weeks
    of the analysis period, 18th June 2021 -- 31st August 2021, whilst the predictive
    distribution is shown for a 2 week prediction horizon with subsequently observed cases
    superimposed.}
\end{figure}

Posterior predictive checking of our model was performed by comparing the 90\% credible interval of the in-sample smoothing distribution of $\tx{I}{R}$ transitions against observed case numbers for the last two weeks of our analysis window. Out-of-sample checking was performed by comparing the 90\% prediction interval of the 2-week-ahead predictive distribution against the subsequent 2 weeks' worth of case data.  These comparisons are plotted for the Lancaster LAD in Figure \ref{fig:predictive-plot}. We see that our model captures both the modest increase and weekly periodicity of the case numbers well, accommodating the apparent ``catch-up'' in cases seen on a Monday after the propensity to test-and-report less at weekends.  In practice, these plots are useful to detecting LADs departing from the expected epidemic trajectory early, so as to alert public health authorities to either case surges or unexpectedly low incidence in individual LADs.

\section{Discussion}
In this paper we have developed an MCMC-based method for fitting a discrete-time spatial
stochastic SEIR model to geolocated case timeseries data for the COVID-19 pandemic in the
UK.  We address particularly the challenge of providing local now-casts and short-term
forecasts of epidemic spread, based on knowledge of population structure and mobility.
Our approach differs from ``classic'' epidemic models \citet[e.g.][]{BirEtAl21}, in that we make no attempt to fit our
model to the entire epidemic timeseries (in the case of COVID-19 from the earliest cases
in February 2020), but choose to analyse only the most recent 12 weeks of data.
This reflects that the fact the epidemic in the UK was highly non-stationary, so that analysing a longer
timeseries would likely have provided little further posterior information and run the risk of being misleading.

Our model represents a trade-off between its complexity and our ability to fit
it to our 380-dimensional timeseries.  It was implemented rapidly in response to the
pandemic, with the innovation being the constrained-space samplers operating on the
censored event data as in Section \ref{sec:censored-data}.  To our knowledge, this is the
first time that such an MCMC scheme has been attempted for a discrete-time model such as
this. 

As a principled fully Bayesian stochastic approach to the modelling and inference of spatio-temporal epidemics, the approach has many advantages. The MCMC algorithm works at scale, allowing the fitting of the COVID-19 model at the LAD level whilst incorporating human mobility in a pragmatic way; we remark that the approach is flexible enough to have been trivially extended to using dynamic mobility data, had it been available.  Importantly, the Bayesian treatment of censored transition event data allows our model to not only provide unbiased parameter inference, but also to provide probabilistic predictions of future case numbers with an improved measure of uncertainty consistent with all sources of modelled noise.  Finally, the sole use of publicly available data and open source hardware/operating system-agnostic computational libraries aids portability and automation of the model.  Indeed, during the COVID-19 pandemic, we deployed the model as an automatic nightly analysis pipeline requiring no manual intervention other than statistical oversight of the results.  Despite the advantages of our approach, however, it is clear that further research is needed to address a number of key limitations as follows.  

With individual-level epidemic models incorporating a high degree of population heterogeneity, it is common to attempt to estimate the baseline $\tx{I}{R}$ transition rate $\gamma_0$ \citep[e.g.][]{JewEtAl09c, McKCookDear09, ChisFer07}.  However, it is known that estimating $\gamma_0$ in the presence of censored event times is problematic, requiring an exceptionally efficient sampler and re-parameterisation of the model to approximately orthogonalise $\gamma_0$ with respect to the censored data \citep{NealRob05}.  With interpretability of the parameters driving our model construction for this application, however, we chose to fix $\gamma_0=\log 0.25$ based on clinically-derived data.  This is a weak assumption compared to other modelling approaches \citep[e.g.][]{epidemia}, though it limits the capability to detect temporal changes in $\gamma_0$ which might occur as a result of improved case detection or changes in the host-pathogen interaction.

Our temporal and spatial random effects, $\bm{\alpha}$ and $\bm{s}$ respectively, were
introduced following the observation that the disease transmission rate appeared to
fluctuate across time and space more than could be explained by human mobility and the
size of the population at risk.  In principle, a spatiotemporal random effect could have been adopted, as is 
commonly used for disease mapping.  However, such
methods increase the dimensionality of the latent surface from $T+M$ to $TM$ and typically
require fast approximate methods to compute the posterior \citep{ScHel11, ZamEtAl12}.
Even if such a method were to be applied to this model, the discrete nature of the
censored event space would still require a Metropolis-Hastings approach such as ours,
though an extremely efficient proposal would be required to ensure an adequate effective sample size.

A further pragmatic assumption adopted by our method is that the case timeseries is a
perfect observation of the number of $\tx{I}{R}$ events per day.  Though this assumption
might hold incontrovertibly for highly pathogenic disease in small populations (such as
foot-and-mouth disease in cattle), for large human populations in which social and
demographic factors affect the propensity of an individual to take a test, a stochastic
observation process would be preferable.  Empirical testing showed that whilst our method
could be implemented if the $\tx{I}{R}$ events were treated as latent -- but informed by
for example a Binomial model -- the resulting algorithm was extremely slow to converge due
to the extra censored data and the interactions within the epidemic event space this
engendered.

The corollary to these limitations is that although our approach
certainly has utility in informing disease control policy during an outbreak, the field of
epidemic modelling is in urgent need of improved methods for inference in the presence of
high-dimensional, correlated, and discrete censored data.  Whilst recent developments in
particle filtering methods have offered promise for improving inference in
individual-level models (e.g. \citep{JuHenJac21, RimJewFear23}), dependent stochastic
metapopulation models such as ours still present a challenge for high-dimensional
importance sampling.  In MCMC methodology, recent advances in non-centering to
orthogonalise state transition events with respect to the model parameters, these are
based on individual-level continuous time models which fail to scale to national-level
human population sizes \citep{PoolEtAl15, PoolEtAl19}.

Finally, all modelling approaches are contingent on reliable and timely data to maximise
their utility in prediction and policy advice.  In this study, we based our analysis on
publicly available case incidence data which as argued above is fraught with uncertainty
due to changes in individuals' preferences towards testing uptake.  However, unbiased
prevalence sampling data was collected in the UK during the pandemic, though was
unavailable at the LAD-level due to concerns over data privacy \citep{EalEtAl22,
  HousEtAl22}.  Nevertheless, the ability to use prevalence \emph{as well as} incidence
data for inference conditional on an epidemic model offers the possibility of improving
parameter identifiability through observations not only of the 1st-order process
(transitions), but also of the 0th-order process (epidemic compartments).  Additionally,
our approach would have been greatly improved with the addition of real-time human
mobility data obtained indirectly through methods such as geolocation via cellular
telephony.  Such telephony data has been shown to be highly effective as a source of
covariate data in epidemic models \citep{GrEtAl20}.  As it stands, our model is capable of
explaining spatial patterns of disease spread via established commuting routes, but cannot
separate spatially-homogeneous variation in human mobility due to social determinants
(holiday periods, media announcements, etc) from the underlying transmissibility of the
virus -- the inclusion of accurate time-varying mobility data in place of our fixed
commuting data would offer the opportunity to surmount this limitation.  We therefore
conclude by recommending that due consideration be made to appropriate sharing of these
data during a future outbreak emergency, so that the full potential of spatial epidemic
models be realised as a resource for policy information and evidence.

\section*{Competing interests}
The authors declare no competing interests.

\section*{Acknowledgments}
ACH and CPJ were supported
through the Wellcome Trust `GEM: translational software for outbreak analysis'
(grant number UNS73114). ACH, CPJ, JR were supported by MRC through the JUNIPER
modelling consortium (grant number MR/V038613/1).  CPJ and GOR were supported by EPSRC (grant number EP/V042866/1).

We are indebted to the Google Research team, who helped us to implement our model, and
fixed bugs in the underlying software libraries swiftly and effectively.

The authors would like to thank The High End Computing facility at Lancaster University
for providing the facilities required for fitting the models in this paper.

The views expressed in this paper are those of the authors and not necessarily those of
their respective funders or institutions.

\section*{Data availability}
All data used in this analysis were public at the time of publication.  COVID-19 case data is available from the UK Government Coronavirus website \citep{coronaviruswebsite, coviddatadownload}.  Snapshots of the covariate data used for the analysis may be found in the accompanying software archive at \url{https://doi.org/10.5281/zenodo.7715657}.

\bibliographystyle{rss}
\bibliography{references}

\end{document}